# TerraServer SAN-Cluster Architecture and Operations Experience

Tom Barclay

Jim Gray

Microsoft Research

Advanced Technology Division

Microsoft Corporation

One Microsoft Way

Redmond, WA  98052

Table of Contents

# TerraServer SAN-Cluster Architecture and Operations Experience

Tom Barclay. Jim Gray ;
{TBarclay, Gray }@microsoft.com

Microsoft Research, 455 Market St., Suite 1690,San Francisco, CA 94105
http://research.microsoft.com/barc

### *Abstract*

Microsoft® TerraServer displays aerial, satellite, and topographic images of the earth in a SQL database available via the Internet. It is one of the most popular online atlases, presenting seventeen terabytes of image data from the United States Geological Survey (USGS). Initially deployed in 1998, the system demonstrated the scalability of PC hardware and software – Windows and SQL Server – on a single, mainframe-class processor. In September 2000, the back-end database application was migrated to 4-node active/passive cluster connected to an 18 terabyte Storage Area Network (SAN). The new configuration was designed to achieve 99.99% availability for the back-end application. This paper describes the hardware and software components of the TerraServer Cluster and SAN, and describes our experience in configuring and operating this system for three years. Not surprisingly, the hardware and architecture delivered better than four-9's of availability, but operations mistakes delivered three-9's.

## 1 TerraServer History: 1998-2003

The TerraServer is one of the largest public repositories of high-resolution aerial image and topographic data. It is designed to be accessed by thousands of simultaneous users using Internet protocols via standard web browsers. During a typical day, TerraServer has about fifty thousand unique visitors who view a million web pages containing 2 to 12 image "tiles". The typical daily peak rate is 40 page-views per second and 225 tiles per second. On peak days we experience 2.5 times the typical traffic.

An image "tile" is a 200x200 pixel Jpeg or GIF file that is stored within a column of a database table managed by Microsoft SQL Server 2000. TerraServer contains 331 million tiles extracted from 260,000 USGS aerial imagery files known as DOQQs (Digital Ortho Quarter Quadrangles) and from 60,000 USGS topographic maps known as DRGs (Digital Raster Graphics.) The USGS provided the 17 terabytes of source imagery as uncompressed TIFF and custom designed graphics files. There is substantial redundancy and overlap within the data files. When compressed, the imagery and necessary meta-data consumes 3.3 terabytes within the SQL database.

Users access TerraServer via the Internet. An HTML-based consumer-oriented interactive application was built using early versions Microsoft Active Server Pages (ASP) and Visual Basic Scripting (VBS) web technologies accessing a single, large database supported by Microsoft SQL Server 7.0 [Barclay98]. This taught us a lot about geography, remote sensing data (imagery), and in particular how to add new data while users were accessing the database.

TerraServer launched in June 1998 with .75 terabytes of compressed USGS aerial imagery covering approximately 30% of the conterminous United States and .25 terabyte of compressed, de-classified Russian military data known as SPIN-2 imagery covering parts of the United States, Europe, and Asia. As we received new data from both sources, we found it very difficult to update the database while users were accessing the system. These update difficulties stemmed from design mistakes in our image processing and load pipeline. This led to a major change in the database schema and web application design that simplified and automated the continuous loading of new data while users were online, without requiring previously loaded data to be available in an uncompressed state [Barclay99].

During the first year of operations, we learned that users typically viewed data of their geographic area and typically used the system between 8 AM to 6 PM in their time zone. Because we had world-wide coverage, the TerraServer web site remained busy 24 hours per day with slight lulls as the sun passed over the major oceans. Thus, all maintenance activity such as database backups, software upgrades, etc. needed to be performed while the system was online.

The 24 hours by 7 days of the week online availability window was an operational challenge for our single large database server hardware architecture. The 2.5 terabyte Digital Equipment Corporation AlphaServer™ had the storage capacity and performance to simultaneously handle the end-user web application, database loading, online database backup to magnetic tape and other online database maintenance activities. However, we were not able to perform system software upgrades, firmware upgrades, or hardware maintenance without shutting down the system and taking the application off-line. There was also no failover scenario should a failure disable the server – recovery would have taken days in the event of a serious hardware failure.

By the end of 1999 the 2.5 TB AlphaServer was completely filled with imagery. It was not practical to expand its storage capacity because of space constraints in the data center. Working with Compaq Computer Corporation, the Windows Server team, and the SQL Server team, we designed a new hardware environment for TerraServer that would implement a highly available, 99.99% or better, environment that would also allow us to perform the following tasks without taking the database or web site off-line:

1. Perform hardware upgrades online – add disk, memory, or tape capacity
2. Perform software upgrades online – operating system patches, SQL Server patches, etc.
3. Tolerate any single hardware failure – disks, processors, memory, etc.

We were motivated to achieve 99.99% availability or higher because the TerraServer user community was changing from a casual consumer users to business use in a wide range of industries. Businesses users complained quickly anytime TerraServer was unavailable. Business users also requested additional capabilities in the TerraServer web application that could only be met with a programmable web services interface. In September 2000, a .NET based Web Service was added to provide a programmable interface to TerraServer meta-data via SOAP/XML protocols. Major users such as the U.S. Department of Agriculture built applications that depended on the new TerraServer Web Service and Map Service [Barclay02].

Microsoft and TerraServer partners such as the USGS are active members of the OpenGIS Consortium (OGC). The OGC promotes an open Web Map Server specification that enables client applications to access and integrate one-or-more OGC compliant Web Map Servers over the Internet. TerraServer added support for the OGC Web Map Server 1.1.1 specification. Later, re-projection services were added to the TerraServer OGC Web Map Server [Jain03]. The re-projection service is used by several public mapping applications including the U.S. National Map program [NATLMAP03] and the Geo-Spatial One-Stop program [GEOSTOP03].

This paper describes the design and architecture of the hardware, system software, and application software involved in the Scalable and Highly Available components of the TerraServer System. We refer to this platform as the TerraServer Cluster and SAN. We operated this configuration from October 2000 through October 2003. Another paper will describe the next generation software and hardware architecture.

# 2 TerraServer Architecture

We wanted the TerraServer database system and web application to be 99.99% available to end users accessing the system via the Internet. Our architecture assumes that the connection to the Internet is 100% available and never fails. In practice the Internet only delivers about 99% availability to end users, but there is little we can do about that in the TerraServer design. To go beyond the four-9's architecture would require a geoplexed solution to tolerate earthquakes, fires, and network outages up-stream from the TerraServer installation. The next generation TerraServer addresses that issue.

The TerraServer high-availability architecture was designed to tolerate and recover from major failures within minutes and from minor failures instantly. It was designed to tolerate the following combined failures:

1. Tolerate the failure of half the database server processing capacity including the failure of any single component such as CPU, memory, disk interface, network interface, etc.
2. Tolerate the failure of half the web application server processing capacity including the failure of any single component such as CPU, memory, disk interface, network interface, etc.
3. Tolerate the single or dual failure of any single device within a disk volume without any interruption to the database software or web application.
4. Tolerate the failure or loss of half the local area network capacity.
5. Enable firmware and software to be patched, upgraded, or expanded online without interrupting service.
6. Enable additional computing capacity, either database or web servers, to be added without interrupting service.
7. Enable online backups to be stored off-site for disaster recovery purposes.

In addition, all hardware and system software failures should be *transparent* to the application software, i.e. no programming changes to the web application were required to achieve the availability goals. The design needed to allow partitioning the data into multiple database servers in order to balance the load across multiple systems. The platform needed to support Windows 2000 and Microsoft SQL Server 2000 which are the basis of the TerraServer web, web service and database applications.

A solution meeting these requirements was designed in the spring of 2000. At that time, two technologies were emerging in the Windows-Intel environment that addressed the above requirements: Storage Area Networks (SANs) and Windows Server 2000 four-node active/passive clustering. Those were the foundation technologies of the TerraServer high-availability design.

## 2.1 TerraServer Cluster-SAN

Many hardware and software vendors implement clustering to address scalability and availability requirements [Hint]. Beginning with Windows NT 4.0, Microsoft offered two-node clustering capabilities for high availability. Sharable disk resources on one computer could logically fail over to a second node should the first node fail. Combined with RAID-1, RAID-5, and RAID-10 storage solutions, clustering provided high-availability and fast-failover for Windows environments. However, two-node clustering did not address scaling beyond a two nodes. Shared SCSI strings were the typical inter-connect method between cluster nodes. Shared SCSI strings have several management, configuration, scalability, and reliability problems that make them unattractive for large-scale solutions.

Windows 2000 Data Center Edition expanded the number of nodes in a cluster from two to four. Configurations could be active-passive (hot-standby), active-active, or other combinations. Storage Area Networks became the preferred processor-storage inter-connection mechanism because they simplified the wiring between nodes and disk resources and had good management tools.

Figure 1 is a block diagram of the TerraServer SAN and Cluster installation. Shaded areas identify the TerraServer SAN and the 4 nodes that form the cluster managed by the Microsoft Clustering Services. Three Extreme Networks Ethernet switches implement the front end network that enables the 10 web servers to access the SQL Server instances running on the Cluster.

We worked with Compaq Computer Corporation to build a Storage Area Network that could accommodate our requirements for several reasons. We had used Compaq StorageWorks™ disk sub-systems on the AlphaServer TerraServer [Barclay98, Barclay99] so we knew StorageWorks™ was very reliable and had excellent performance. Tests of SQL Server 2000 and Windows 2000 Server supporting the USGS topographic data-set on a Compaq ProLiant 8500 8-way Intel processor performed better than the DEC AlphaServer. The new server was as reliable and fit in 1/8th the rack space.

The TerraServer initially used three Compaq StorageWorks™ Enterprise Storage Arrays 10000 (ESA-10000). A single ESA-10000 occupies a 42u computer rack. Each ESA-10000 was sub-divided into 3 sections. Each section contained four shelves. One shelf contained a redundant pair of HSG-80 disk controllers. The other three shelves housed the SCSI disks that were connected to both HSG80 controllers on the first shelf. The top section of each ESA-10000 rack contained eighteen 18GB, 15kRPM SCSI disk drives. Meta-data and other frequently accessed tabular data were stored on these drives for fast access. The bottom two sections of each ESA-10000 rack contained sixty 73GB, 10kRPM SCSI disk drives. The Imagery tile data was stored on these drives (see Figure 2).

Figure 1: TerraServer SAN Cluster Block Diagram

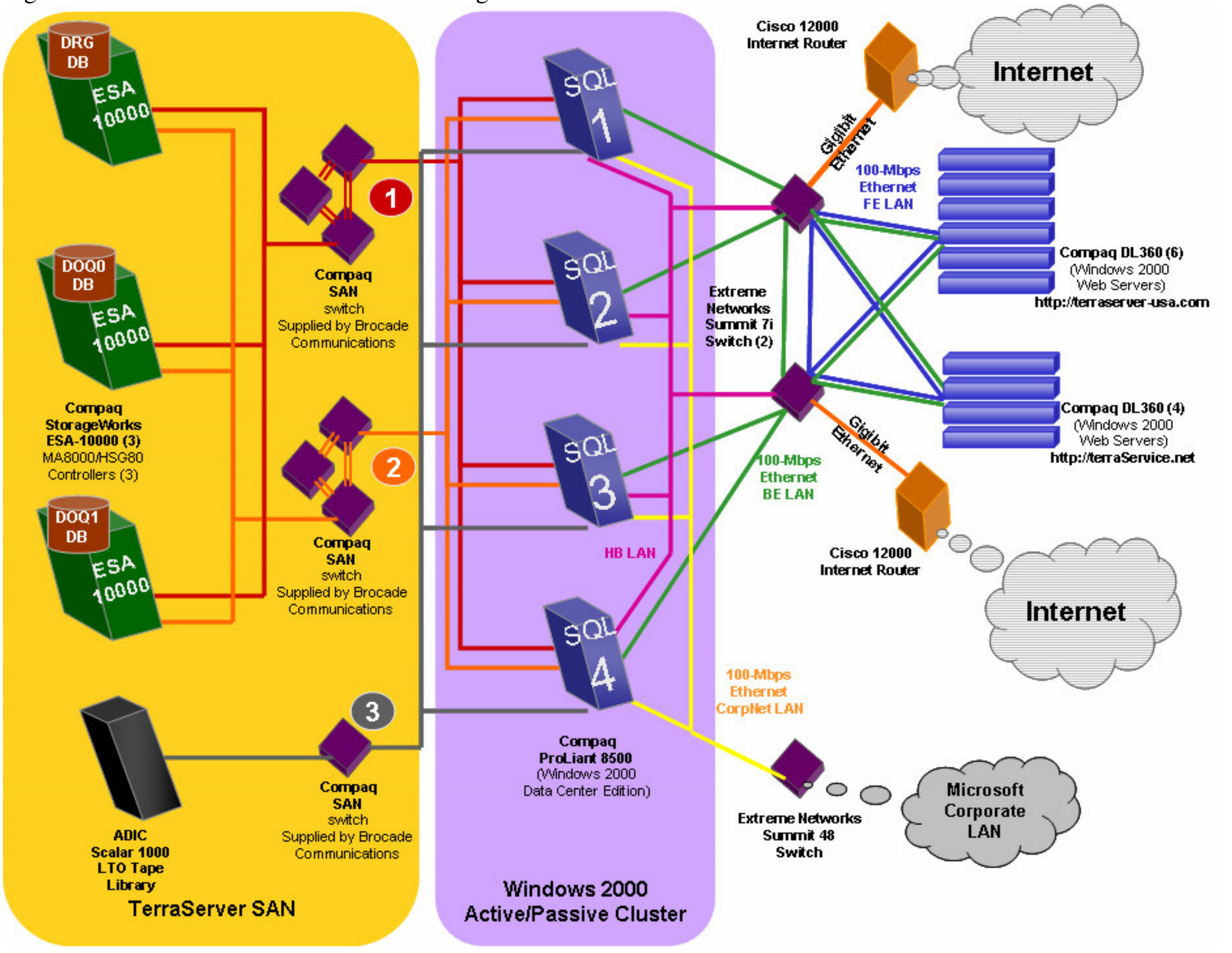

HSG-80 disk controllers implement RAID 0, 1, 5, or 10. We chose to implement RAID-10 – stripes of mirrors – for TerraServer disk volumes. Three physical disks were used to store the contents of one logical volume (triple mirrored). Thus, two disks must fail in a single mirror set to cause data unavailability. We never saw that happen – we saw about 24 disk failures but they were all independent.

## 2.2 SAN Disk Configuration

Each of the three TerraServer databases stored on the TerraServer SAN has the same logical and physical schema. The majority of the bytes in each database are stored in the Imagery table in a blob (binary large object) column. This table can exceed one terabyte per database. The other TerraServer tables contain tabular data– integer, string, datetime, floating point values. Most rows are in these meta-data tables; however, the sum of the meta-data tables is less than 100 gigabytes.

We placed all meta-data tables on the faster-smaller, 15k RPM 18GB disks because these tables had more traffic (were hotter). The Imagery table was placed on the slower-larger, 10k RPM 73 GB disks. The top sections of the first three ESA-10000 arrays were dedicated to the 15k RPM, 18 GB hard drives. The lower two sections of the first three ESA-10000 arrays were dedicated to the 10k RPM, 73 GB hard drives. Because TerraServer has less than 100 GB of meta-data, the top section was not fully populated with 18 GB hard drives. The lower sections were fully populated with 73GB hard drives.

Each logical disk was represented as three physical disks. For the high-speed disks, the RAID controllers presented a stripe of two such triple-mirrors as a single 36GB volume. Then three of these volumes were striped together using controller-based raid to present a single 102 GB volume to the Windows cluster. Five triple-mirrored volumes of the slower speed, 10k RPM large drives were striped using controller-based raid to present a single 339 GB volume to the Windows cluster.

Initially the TerraServer SAN was configured with three ESA-10000 storage racks. The SAN presented a total of three high-speed, 102 GB volumes, and 12 lower-speed, 339 GB volumes to the Windows cluster. In the fall of 2001, we added a fourth ESA-10000 populated entirely with 90 10k RPM 73 GB SCSI disk drives. Instead of striping five tripled-mirrored volumes, we striped ten tripled-mirrored volumes and presented three, 678 GB volumes to the Windows cluster.

The HSG-80 disk controllers were connected to the host computers via 1Gbps fiber channel connections managed by Brocade Silkworm-2800 SAN switches. For redundancy, there were two separate SAN networks, called *fabrics*, which connected the disk controllers to the host servers. The fabrics are labeled ❶ and ❷ in Figure 1. Each HSG-80 controller and each host server was connected to both fabrics. I/O requests from a host server and data returned from an HSG-80 controller could pass over either fabric. The SAN switched automatically reroute requests over an available fabric should a SAN switch or fabric fail.

Each SAN fabric had a total of 28 end points – 4 Host Bus Adapters (HBA), one per database server, and 24 HSG-80 connections, 2 per ESA-10000 section, 3 sections per ESA-10000, and 4 ESA-10000 (2 x 3 x 4 = 24). A Brocade Silkworm-2800 SAN Switch has a total of 16 ports. Thus each SAN fabric needed three Brocade Silkworm-2800 switches to route traffic between all HSG-80 controllers and all four hosts in the cluster.

A seventh Brocade Silkworm 2800 implemented a third fabric, labeled ❸ in Figure 1, to support tape backup and restore operations. We used an ADIC Scalar 1000 ™ tape library containing 4 LTO (Linear Tape Open) tape drives. The SQL databases were regularly backed up to tape using Veritas NetBackup™ software integrated with SQL Server online database backup utility. Tapes were removed from the tape library and stored off-site.

## 2.3 File System Design

It is possible within the Microsoft Cluster environment for all the disk devices to be owned by a single node within the cluster so all device names must be unique within the cluster. SQL can use Windows *device numbers* and

Figure 2: TerraServer Installation Photograph

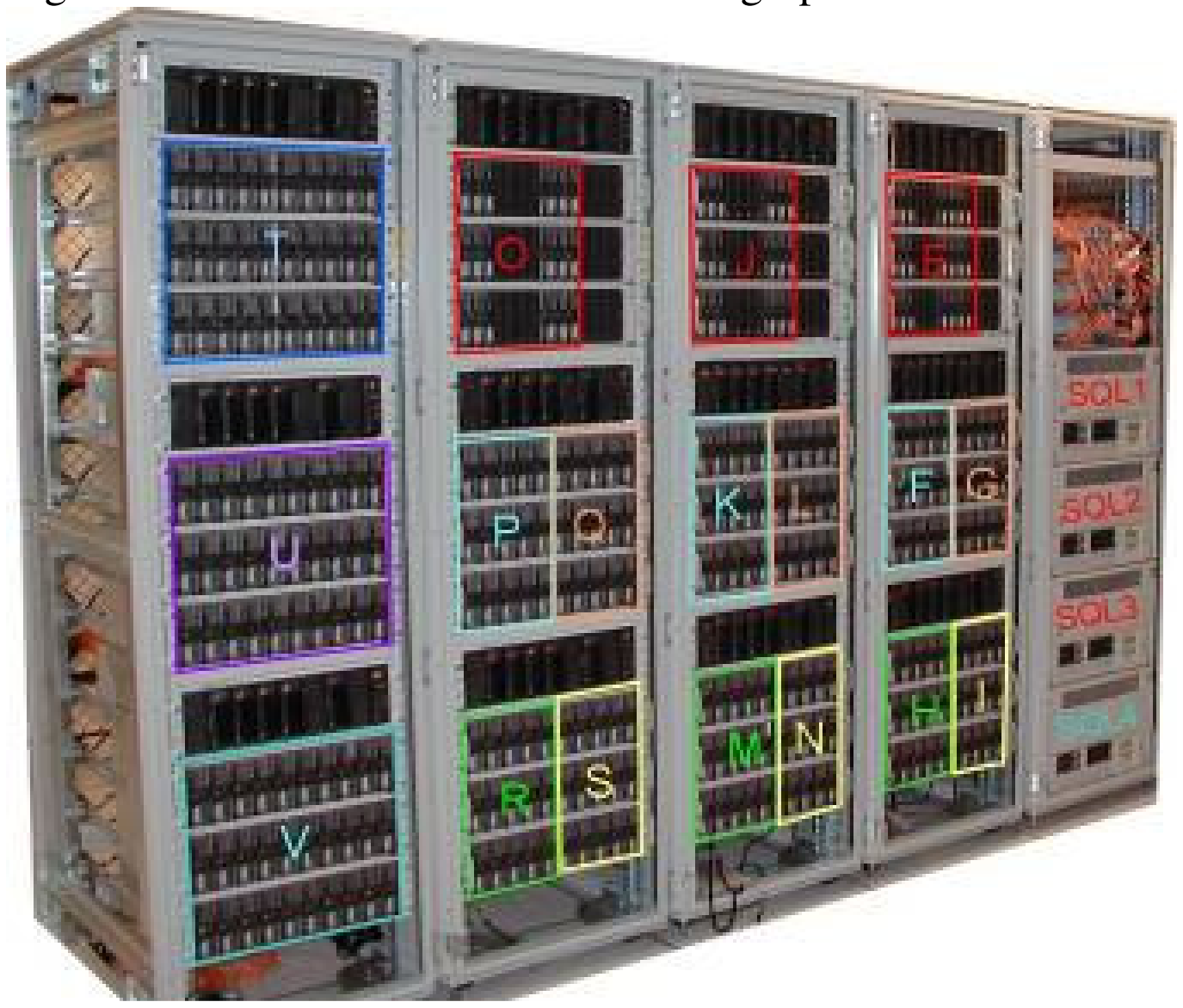

*mount points* to name file systems, but the TerraServer application was able to fit within the E: through Y: drive-letter space.

Figure 2 is a photograph of the TerraServer Installation overlaid with the Windows file system volume assignments to the three ESA-10000 sections. Drives E:, F:, G:, H:, I:, and T: contained the database files belonging to the DRG_DB resource groups and the DRG SQL database. Drives J:, K:, L:, M:, N:, and U: contained the files belonging to the DOQ_0 resource group and the DOQ0 SQL database. Drives O:, P:, Q:, R:, S: and V: contained the files belonging to the DOQ_1 resource group and the DOQ1 SQL database.

Drives E:, J:, and O: stored the meta-data used by the web pages that formed HTML web pages. All other drives stored the imagery data. The meta-data is the most frequently accessed data. Some queries such as the *search for imagery by place name* and s*earch for imagery by geographic coordinate* require joins of several tables and range scans of the B-Tree indices which can generate many physical database I/Os. Thus the meta-data was placed on the higher-speed disk volumes.

## 2.4 Database Partitioning

We chose an active-active-active-passive clustering architecture to handle the occasional spikes in traffic to the system. Since there were three active servers, the TerraServer database was partitioned into three resource groups. Usage patterns explained below suggest a natural database partitioning Shown in Figure 3.

The TerraServer has two major imagery data-sets (1) one-meter resolution USGS Aerial Imagery (a.k.a. USGS DOQ or DOQQ) and (2) two-meter resolution USGS topographic maps (a.k.a. USGS DRG). The DRG data-set occupies 900 gigabytes and the DOQ occupies 2.5 terabytes.

Users access the DOQ data more frequently because the data is fresher, larger, and rarer (several web sites host topographic data whereas few host USGS DOQ data).

Normally, TerraServer is the busiest on Monday and Tuesday averaging 50k unique visitors accessing 1.2 million web pages and accessing 5 to 6 million image tiles per day. Weekends have the least volume averaging 30k unique visitors accessing 700k web pages and 4 million tiles per day. Occasionally, TerraServer is referenced in the popular press (e.g., television, press, or website). When this occurs, it is not unusual to see three times the normal traffic. The day of one such event, TerraServer had 277 thousand unique visitors, viewing 4.5 million pages, and 12 million tiles.

Figure 3: TerraServer Database Partitioning

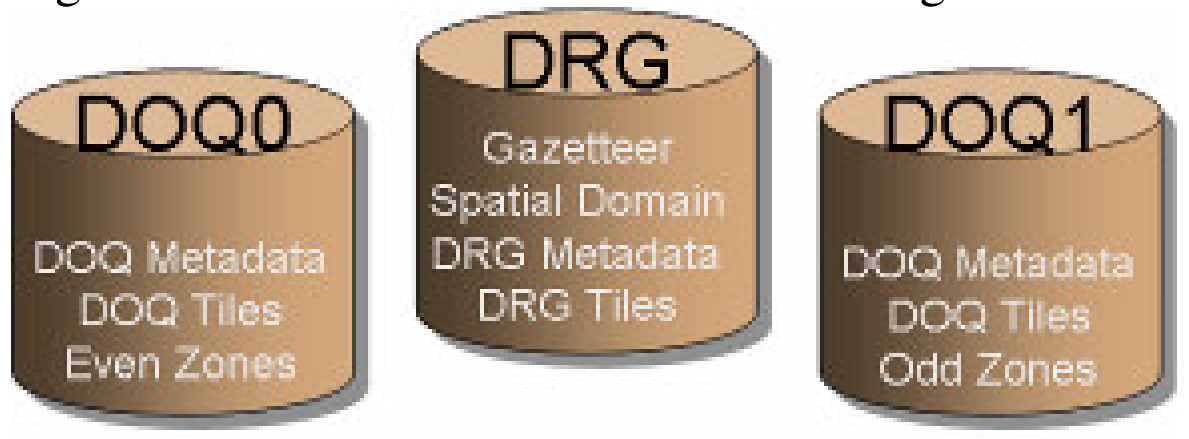

We observed that users had a tendency to visit the site at the beginning of the work day in their time zone and view imagery of locations near where they live. The disk subsystem is initially busy reading data for a single time zone into memory. Disk traffic slows down as additional users from the same time zone access data already cached in memory. When the users from the next time zone arrive at work, the disk subsystem becomes busy reading imagery for the next time zone into memory; then it quiets down as new users in that time zone accessed data already fetched from disk. So, it is natural to partition the database by time zones and to spread the zones among the servers. We used this geographical and temporal locality to load-balance the system.

The USGS data is organized into Universal Transverse Mercator zones (UTM zones). A single zone is a North-South stripe about 6 degrees wide (about 350 miles of 500 km). The conterminous United States touches ten UTM zones (10 thru 19) and four time zones (Pacific, Mountain, Central and Eastern).

The TerraServer web application's database accesses can be divided into three broad categories [Barclay99]:

a. Scan the Gazetteer and Imagery spatial tables looking for candidate imagery.
b. Scan the Imagery meta-data and Imagery spatial tables to fetch the data required to build an HTML web page referencing imagery and other TerraServer pages.
c. Fetch one or more image tiles

Putting all these facts together suggests the following partitioning design. The Gazetteer is replicated in all three servers. The smaller and lower traffic USGS DRG data (topographic maps) were placed on one database server, and the USGS DOQ data (photos) was UTM-zone interleaved between two database servers. The odd UTM zones were placed together on one database server. The even UTM zones were placed on a second database server. Effectively, this split a single time zone across three database servers.

This database partitioning spread the I/O load fairly evenly across all three active database servers.

## 2.5 Cluster Resource Groups

The Microsoft Cluster Services (MSCS) and SQL Server 2000 software achieve failover transparency by presenting a logical server and database naming system (resource group name) that is independent of physical system names. In a non-clustered environment, applications access a SQL database by specifying the system name and database name. The client connection software uses the TCP/IP network APIs to find the computer system and route a connection request to the SQL Server software running on that node.

In a clustered environment, applications pass a resource group name that includes the named instance of the SQL environment to connect to. The MSCS cluster software resolves this name to an IP address that maps to the cluster node that currently owns (hosts) that resource group. When the request reaches the physical server, the SQL instance name in the connection request is used to route the request to the correct SQL Server software process handing the database the client application wants to access.

The notion of a "named SQL instance" was a new feature of SQL Server 2000. In prior versions, only a single copy of the SQL Server system could run on a server at a time. The SQL Server system could handle requests for multiple databases simultaneously. However, the database applications would have to run within a SQL environment configured the same for all databases. With SQL Server 2000, multiple copies of the SQL Server system can execute simultaneously on a single server. This allows a set of databases "A" to run with the SQL Server system tailored to their needs, and another set of databases "B" to run with a SQL Server system configured differently. Also, the separate SQL Server systems, a.k.a. an instance, can be paused, stopped, and started separately without affecting any other instances on the system.

The MSCS system uses the named instance feature to transfer database resources from one node to another in the cluster. If a server node supporting a named instance fails, the MSCS system automatically transfers the named instance to a healthy node in the cluster. The SQL Server system begins execution on the new node by recovering any databases that have in-flight transactions in progress. When recovery is complete, the SQL Server system will begin to accept client connection requests.

## 2.6 Cluster Administration

The four servers, labeled SQL 1, SQL 2, SQL 3, and SQL 4 in Figure 1, formed a 3-active 1-passive (hot standby) cluster. The Microsoft Cluster Services [MSCS03] of Windows 2000 Data Center Edition implemented the cluster services. SQL Server 2000 Enterprise Edition integrates with Microsoft Clustering Service. Microsoft Cluster Services supports a concept of *resources* and *resource groups*. A resource is a hardware or software entity than can be migrated across nodes in the cluster. Disk devices, IP-addresses, databases, and applications are typical resources. Shareable resources are grouped together as a resource-group and given a logical name and a list of nodes which can own the group. All resources in the group failover together when a group moves to a new node. At any instant, a single node *owns* a group of resources.

The TerraServer cluster was named TkTerraClstr. TkTerraClstr contained four resource groups – Cluster Group, DRG_DB, DOQ_0, and DOQ_1. The Cluster Group contained resources central to the cluster service such as Cluster Name, Cluster IP, and other resources. The resources in the DRG_DB, DOQ_0, and DOQ_1 were specific to one instance of a SQL Server database. The resources included the disk volumes, SQL instance's name, IP address, and other information specific to SQL Server.

Each physical node in a cluster may own one or more resource groups at any time. The default configuration was for the SQL-1 node to own the DRG_DB resource group, the SQL-2 node to own the DOQ_0 resource group, and for SQL-3 node to own the DOQ_1 resource group. SQL-4 was the passive node. The administrator could use the Microsoft Cluster Service administrator utility to change resource ownership. The utility provides a GUI interface that allows the system administrator to move a resource group from on node to another. This procedure is effectively what the cluster service automatically does if a node fails. The TerraServer SAN experienced a 30 to 45 second outage for that database partition when a resource was taken off-line, moved to a new node, and restarted.

## 2.7 Local Area Network Architecture

The TerraServer SAN had four virtual local area networks (VLAN) (see Figure 1).

1. The HeartBeat LAN, depicted in magenta and labeled HBLAN, was used by Cluster Service to verify that nodes were operational. If communication was lost to a node, then the cluster service automatically moved cluster resources to surviving nodes.
2. The Back End LAN, depicted in green and labeled BELAN, connected the web servers to the database servers. The ASP.NET web page classes used ADO.NET to invoke SQL stored procedures to read data from one of the three TerraServer SQL databases.
3. The Front End LAN, depicted in blue and labeled FELAN, connected the web server nodes to the Internet. This isolated the database servers on the Back End LAN from the Internet and so was an addition firewall.
4. The CorpNet LAN, depicted in yellow, connected the database server nodes to the Microsoft Corporate LAN. This enabled the databases to be loaded from servers located on the main Microsoft campus. In addition, the database servers could be remotely managed using Microsoft Terminal Services and other remote management tools such as SQL Enterprise Manager. The CorpNet LAN connection was not redundant. However, the web application could operate without the CorpNet LAN connection being active.

All four LANs were implemented by the Extreme Network Summit Gigabit network switches. Two Summit 7i switches implement the Back End and Front End LAN. Half the web servers and two of the database servers were connected to one Summit 7i switch. The other half of the web servers and other two database servers were connected to the second Summit 7i switch. Each Summit 7i switch had an up-link to the Cisco 12000 routers that provide the connection to the Internet. Physically, the Cisco 12000 routers were in different location within the data center and are owned and operated by the MSN network group. They provided a firewall between the TerraServer and the Internet. A third switch, a Summit 48, implemented the HeartBeat LAN and the CorpNet LAN.

## *2.8 Web Server/Service Architecture*

The TerraServer web application has a client-server architecture. Web Servers host the web page generation software, implemented as ASP.NET programmable web pages. The ASP.NET page classes use ADO.NET to connect to the TerraServer databases, execute a stored procedure, and translate the returned data to HTML or an image mime type. The ADO.NET SQL connection is made using the virtual name of the SQL database and virtual IP address. This allows the database instance to move from one physical node to another within the cluster without any change to the client code.

There are two TerraServer web sites. http://terraserver-usa.com implements the interactive human HTML-based interface, and http://terraservice.net implements the SOAP/XML based programmatic interface typically used by professional programmers who embed the TerraService in other applications. Several web servers were deployed for each web site so that each site is fault-tolerant and has enough processing power. To meet the availability requirements, we configured twice the required number of servers needed to support the usage load on each web site. This ensured that we had the needed capacity when a web site got very popular or if one or more web server nodes failed. We configured six web servers to support the interactive web site and four web servers to support the SOAP/XML web service. These servers are very inexpensive so we could afford to overprovision them.

## *2.9 Load Balancing the Web Servers*

We explained how database servers failover and how the load is balanced among them. We used the Symmetric Load Balancing (SLB) feature of the Extreme Network's switches to balance the load among the web servers. The SLB feature automatically detects active web servers. We could add and subtract web servers and change their software and the Extreme Networks equipment would automatically route around the down servers and reintegrate them when they returned to service.

The TerraServer SAN configuration supported two web sites http://terraserver.microsoft.com (later known as http://terraserver-usa.com) and http://terraservice.net. Six web servers supported the http://terraserver.microsoft.com web site and four web servers supported the http://terraservice.net web site. The web servers for each site were evenly split between the two Extreme Networks Summit 7i switches in the TerraServer configuration.

The Symmetric Load Balancing feature operates within a single Summit 7i switch. If one of the servers supporting http://terraserver.microsoft.com failed, then the Summit 7i automatically and transparently routed the request to one of the other available web servers allocated to http://terraserver.microsoft.com. End users only notice a failure if all the web servers allocated to a web site failed. That never happened.

## *2.10 SAN I/O Performance*

The primary design goal of the TerraServer SAN and Cluster configuration is high availability. While not the primary design goal, I/O performance was still very important.

We measure the maximum throughput of the drives and the SAN using synthetic I/O profiles. We used the SQLIO.EXE program originally written by to asses the performance of a disk sub-system running SQL Server. This is the same program and technique we used to measure the performance of Serial ATA drives [BARC03].

SQLIO can be configured to and measure read versus write, and sequential versus random performance varying the number of memory buffers, number overlapped (parallel) I/O operations, etc. We configured the following SQLIO tests:

- Sequential 64KB 4-deep read and write requests.
- Random 8KB read and write using 1, 4, 8, 16 and 64-deep (outstanding) requests.

The tests were executed on the following drive configurations:

- One 15k RPM RAID-10 volume.
- One 10k RPM RAID-10 volume.
- One 15k RPM RAID-10 and 4 10k RPM RAID-10 volumes (disk capacity available to one SQL Server database instance)

Figure 5 shows the read performance. The TerraServer web application performs Random Read I/Os of index and data pages. The Random Read with 4 overlapped I/Os simulate how SQL Server performs I/O operations on behalf of the TerraServer application.

Surprisingly, there was not much difference between the 15k RPM volume and the 10k RPM volume. The difference between a single volume tests and the five volume test shows that the SAN switches and network scaled to approximately two times the single volume test.

Figure 6 shows the read throughput in mega-bytes per second. It appears that the maximum bandwidth though the SAN network is 100 MB per second.

The TerraServer Load System performs a combination of Random Read and Write operations. Figure 7 depicts the write performance: Again there is not a large difference in I/O operations per second between the 15k RPM and 10k RPM drives.

Figure 8 shows the write throughput in MB per second. Both the TerraServer web application and the TerraServer Load System I/O requirements were well within the performance profile of the SAN.

Figure 5: Read I/O Operations Per Second

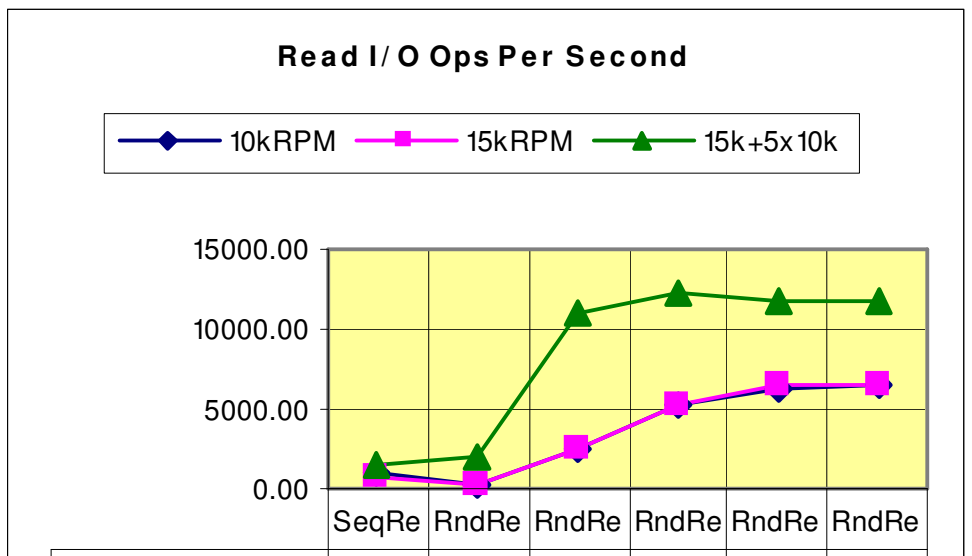

| | SeqRe | RndRe | RndRe | RndRe | RndRe | RndRe |
|---|---|---|---|---|---|---|
| 10kRPM | 885.92 | 360.21 | 2423.0 | 5158.9 | 6374.8 | 6476.0 |
| 15kRPM | 809.51 | 351.24 | 2420.8 | 5193.2 | 6557.2 | 6601.8 |
| 15k+5x10k | 1568.5 | 1953.4 | 11098 | 12140 | 11781 | 11782 |

Figure 6: Read bandwidth (MBps)

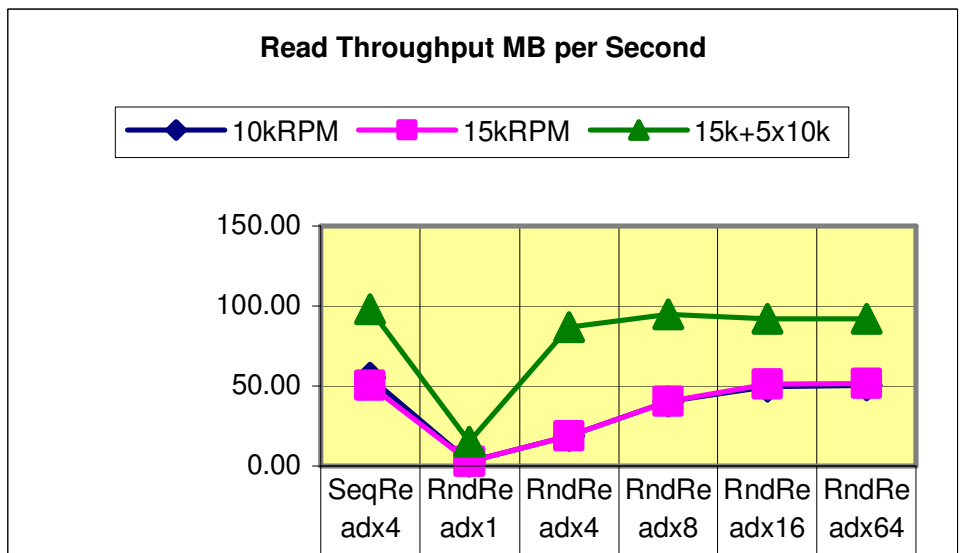

| | SeqRe adx4 | RndRe adx1 | RndRe adx4 | RndRe adx8 | RndRe adx16 | RndRe adx64 |
|---|---|---|---|---|---|---|
| 10kRPM | 55.37 | 2.81 | 18.93 | 40.30 | 49.80 | 50.59 |
| 15kRPM | 50.59 | 2.74 | 18.91 | 40.57 | 51.22 | 51.57 |
| 15k+5x10k | 98.03 | 15.26 | 86.7 | 94.84 | 92.04 | 92.04 |

Figure 7: Write I/O Operations Per Second

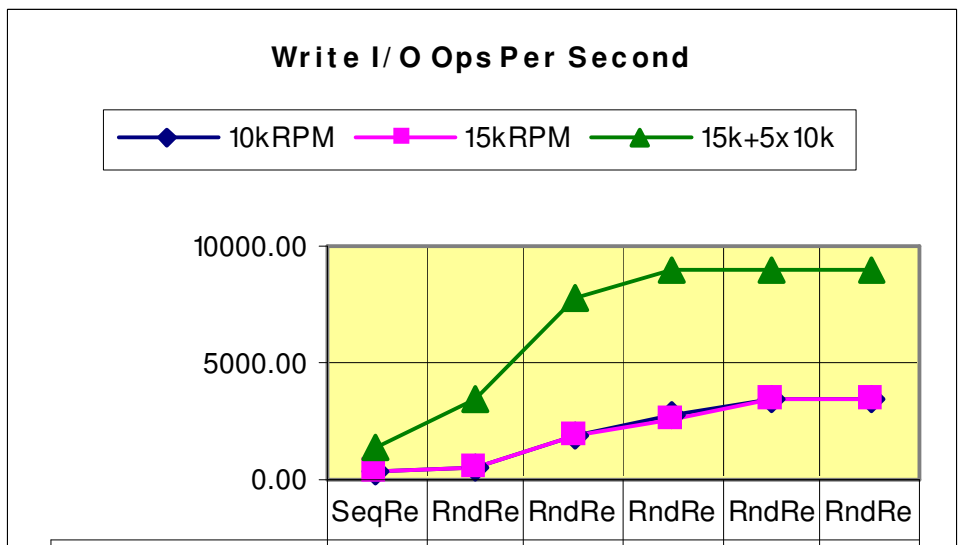

| | SeqRe | RndRe | RndRe | RndRe | RndRe | RndRe |
|---|---|---|---|---|---|---|
| 10kRPM | 349.73 | 500.50 | 1936.6 | 2729.0 | 3397.1 | 3417.3 |
| 15kRPM | 345.69 | 527.61 | 1976.9 | 2600.2 | 3477.0 | 3488.9 |
| 15k+5x10k | 1370.6 | 3445.2 | 7695.8 | 8893.4 | 8887.6 | 8931.7 |

Figure 8: Write Throughput

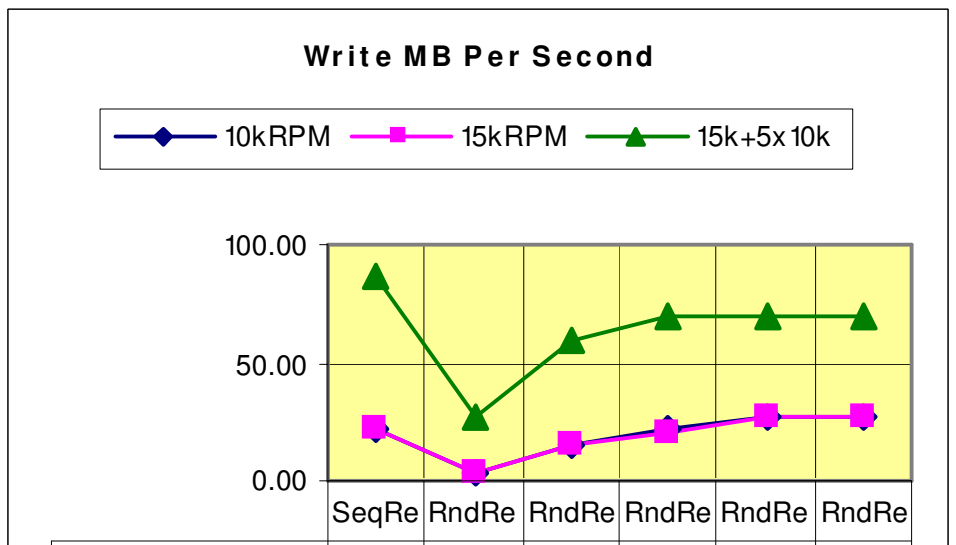

| | SeqRe | RndRe | RndRe | RndRe | RndRe | RndRe |
|---|---|---|---|---|---|---|
| 10kRPM | 21.85 | 3.91 | 15.13 | 21.32 | 26.54 | 26.69 |
| 15kRPM | 21.60 | 4.12 | 15.44 | 20.31 | 27.16 | 27.25 |
| 15k+5x10k | 85.66 | 26.91 | 60.12 | 69.47 | 69.43 | 69.77 |

Both the TerraServer web application and the TerraServer Load System I/O requirements were well within the performance profile of the SAN.

## 2.11 Tape-Based Disaster Recovery

The TerraServer SAN Cluster was a technology demonstration. If it were a real line-of-business application, then a second or third site would have been built in order to survive Internet network outages or natural disasters such as earthquakes and fires. For demonstration purposes, we limited our disaster recovery capability to restoring the databases and applications from tape should a physical disaster or an operations error occur that destroyed our data or the equipment supporting the data.

We wanted to be able to recover the databases from data preserved at off-site locations. We assumed that the hardware configuration survived any physical disaster or was replaced exactly as it was prior to the disaster. Thus we chose a traditional approach of backing the database and our application to tape, and storing the tape media off-site.

The AlphaServer configuration deployed in June 1998 used a StorageTek Timberwolf 9710 tape library containing 10 DLT (Digital Linear Tape) drives. The tape library was directly attached to the AlphaServer. We attempted to use the StorageTek library on the TerraServer SAN and Cluster, but quickly ran into two issues. First, we wanted the tape library to be connected to the SAN instead of connected to a single server to simplify sharing the tape library among the cluster nodes. The Timberwolf 9710 could not connect to the SAN directly. Second, we had difficultly restoring multi-terabyte database backups. There were known issues with reading DLT media on a device different than the device that wrote it. In our particular case we had 10 DLT drives in the library to deliver adequate I/O performance. A terabyte backup consumed more than forty DLT tapes and needed several drives concurrently to complete within eight hours. The probability of a device reading tapes that it did not write was near zero. Frequently, TerraServer restores would fail because the tape library could not successfully read all forty tapes of the backup set. Fortunately, we never actually needed to restore from tape.

We abandoned DLT media after a backup on an eight-drive DLT library and restore on a second vendor's ten-drive DLT library failed. Backing up on one library and restoring on a different library is likely occurrence if a disaster destroyed equipment. We looked for a tape solution that could reliably backup and restore on physically different tape libraries.

There were a number of alternative tape media types available in the spring of 2001. Linear Tape Open (LTO) was attractive for several reasons:

- Drives and media were available from more than one manufacturer (IBM and HP). Both companies have a reputation for high quality products.

- The native (non-compressed) capacity of 100 GB per tape was four times better than DLT-IV (28GB.)
- The LTO transfer speed enabled four drives to backup a terabyte in less than eight hours where it required eight DLT drives to do the same.
- The ADIC Scalar 1000 supported fiber channel connection directly to the TerraServer SAN. The library could be shared between all nodes of the cluster using the SAN instead of the LAN.
- The ADIC Scalar 1000 could store three complete backup "savesets" of each database partition before reusing or replacing tapes.

ADIC provided a Scalar 1000™ that connected directly to the fiber-channel TerraServer SAN [ADIC01]. We used Veritas' NetBackup™ DataCenter utility to manage the Scalar 1000 library [VER01]. NetBackup is integrated with SQL Server 2000 Backup and Restore utility. Backup operations occur online and are initiated by the NetBackup utility.

We performed ten separate backup and restore tests. Each database partition was backed up using a different Compaq server. The tapes were moved from the ADIC 1000 library in the production data center to a second ADIC Scalar 100 with four LTO tape drives SAN connected to a 2 terabyte disk farm. The first restore failed because we had incorrectly configured the Scalar 1000's tape labeling. Once this was corrected, we could reliably move savesets with 10+ LTO tapes and restore them on a different library (9 out of 9 tests passed.)

**Table 1: Tape Backup Performance**

| | |
|---|---|
| Data Bytes Backed Up | 1.05 TB |
| Total Time | 7.9 Hours |
| Number of Tapes Consumed | 12 |
| Total Tape Drives | 4 |
| Data Throughput | 137 GB/Hour |
| Average Throughput Per Device | 34 GB/Hour |

Table 1 shows the typical performance of the Scalar 1000 and NetBackup software backing up one of the three TerraServer databases. The SQL Server backup utility only copies database pages that contain data. Therefore only 1.05 of the 1.2 TB database was actually transferred to tape.

A copy of the NetBackup utility existed on each database server in the cluster. The NetBackup Administrative GUI program controls the scheduling of backup operations. We performed full backup operations on weekends. The DRG database would be automatically backed up on Saturday morning at 2 AM, followed by the DOQ0 database at 2 PM, and the DOQ1 database at 2 AM on Sunday. The DOQ1 database was the largest database and would typically take 8.25 hours to backup 1.33 terabytes of data.

We chose full backups rather than incremental or log backups in order to simplify disaster recovery operations. The TerraServer web application is read-only. The data-loading process was completely in our control. The load programs paused when database backups were scheduled to occur. This reduced the load on the database servers during backup and simplified data recovery. If a disaster occurred, we only had to locate and process one backup save-set. Other more elegant approaches would have required processing the last full backup, and then process each incremental backup in the correct order. Since a data recovery operation is expected to be a very rare event, we chose the simplest recovery method fearing that more complicated approaches might fail. In practice, we never lost any data on the database cluster which required us to perform a database restore.

## 2.12 Architecture Summary

The hardware and system software architecture met our requirements and goals for high availability.

The three active and one passive cluster delivered approximately 100% increase in processing power enabling the application to run comfortably at peak load with three servers or stressed with just two of the four servers. For simplicity, we deployed double the number of web servers required for normal operation. This allowed us to tolerate several web servers being inactive while still supporting our usage spikes.

There was no single point-of-failure between the Internet and the disk storage. Users could reach data of interest via multiple paths starting from the two separate Internet routers, two separate network switches, four database servers, two SAN fabrics, two redundant disk controllers, and three physical disks storing the data.

Usage patterns observed during the two years of operation prior to deploying the TerraServer SAN and Cluster educated us on how to partition the data to achieve a balanced I/O load across the three data partitions. The meta-data had the highest IO traffic and so was stored on the high-speed drives of one database partition. The least popular imagery data was also stored on the slower drives of the partition supporting the meta-data. We redundantly stored the frequently accessed meta-data on the other two partitions in case the one database server was unable to handle the volume. However, the single database server was able to handle all meta-data requests.

The ADIC Scalar 1000 tape library and Veritas' NetBackup DataCenter utility provided a reliable and efficient disaster recovery solution. Database backup of a terabyte database partition was performed within eight hours with minimal impact to the database server. We never had a catastrophic disk failure on the TerraServer SAN that required a database restore. However, we did test and we able to restore our databases in a remote data center using a separate tape library.

The TerraServer SAN and Cluster met all the availability requirements identified at the beginning of Section 2. The following sections describe our experience operating the TerraServer SAN and Cluster from September 2000 through October 2003.

# 3 Operations Experience

We operated the TerraServer SAN and Cluster described in Section 2 for three years. The SAN was first deployed in Microsoft's Tukwila-1 data center in September 2000. It began operational service during the Windows 2000 Data Center Edition launch in San Francisco on September 29, 2000. This section documents our experience operating the SAN and Cluster from its launch until the system was retired in early November 2003.

Our goal was for the database cluster to be 99.99% available, a.k.a. "four nines", as measured from the web servers. End users might experience lower availability due to network-switch and web-server failures in front of the database cluster; or due failures further up stream. Unfortunately we fell short of goal due to operations mistakes discussed later in this section.

The TerraServer SAN and Cluster supported a single logical database physically partitioned into three separate SQL Server databases. The TerraServer web application can deliver *some* imagery if only one or two of the three databases are accessible. We measure downtime as the time that *any* of the three databases was unavailable.

We categorized and tracked downtime as planned or unplanned, and hardware, software, or operations related. The following are how we categorized downtime:

- Hardware error or failure (unplanned)
- Hardware/firmware maintenance (planned)
- System Software (O/S or SQL) failure (unplanned)
- System Software maintenance (planned)
- Operation Errors (unplanned)
- Operation Activity (planned)

The following sections describe four key issues:

- Transparent handling hardware and software failure.
- Online configuration changes.
- Hardware and software fault rates and tolerance.
- Operating a cluster environment.

## 3.1 Failover Transparency

A major requirement of an availability solution is transparency of system component failures to the TerraServer application. TerraServer is a web application that depends on the SQL Server databases to translate user requests into query results that are returned to the user in the form of HTML web pages and graphic images. We did not want an availability solution that would require substantial logic within the application to detect or system component failures.

As explained in Sections 2.5 and 2.6, Microsoft Cluster Services and SQL Server 2000 software present failover transparency by presenting a logical server and database naming system (resource group name) that is independent of physical system names. In a non-clustered environment, applications access a SQL database by specifying the system name and database name. The client connection software uses the TCP/IP network APIs to find the computer system and route a connection request to the SQL Server software running on that node.

In a clustered environment, applications pass a resource group name that includes the named instance of the SQL environment to connect to. The MSCS cluster software resolves this name to an IP address that maps to the cluster node that currently owns (hosts) that resource group. When the request reaches the physical server, the SQL instance name in the connection request is used to route the request to the correct SQL Server software process handing the database the client application wants to access.

The client application cannot tell which physical server is servicing requests. The client is only aware of the logical system and SQL Server instance name. When a failover event occurs, the in-flight operations of over open connections terminate with a network error. New connection requests receive the equivalent of a *server not found* error while the database resources are being moved to a new server.

The only change to the TerraServer application required to run in a clustered environment was the name of the server to connect to. The Active Data Object (ADO and ADO.NET) component was used by the TerraServer application to communicate with the SQL Server system. The connection method requires a text string containing name and value pairs that ADO.NET uses to communicate with the correct SQL Server instance. A typical connection string on the cluster looks like:

```
Server=terravs1\inst1;Database=ts4Drg;
Integrated Security=true;Network=dbmssocn;
```

A connection string in a non-clustered looks like:

```
Server=terraserver;Database=ts3;
Integrated Security=true;Network=dbmssocn;
```

All connection strings in the TerraServer application are maintained in external configuration files (not hard-coded within the application.) Thus, no code changes were required to port the application from a non-clustered environment to a clustered environment.

We also chose not to add special-case code in the application to recognize a failover event. At the application level, there isn't any information that differentiates a temporary unavailability while resources are failing over from a catastrophic failure that the system cannot recover from. To the end user, a failure within the cluster looks no different than an intermittent network error. Web users are conditioned to refresh their browser to correct intermittent Internet errors. Since resource failover operations completed within 30 to 45 seconds, we decided not to add any special error handling within the application and depended on the user to refresh to reconnect.

## 3.2 Cluster Expansion

The configuration depicted in Figure 1 is the final configuration. Initially the SAN and Cluster was configured with:

- three ESA-10000 storage racks instead of four,
- two SAN fabrics instead of three, and
- a direct-attached StorageTek Timberwolf tape library not a SAN-attached ADIC Scalar 1000 tape library.

A clear benefit of the SAN and Cluster configuration is the ability to expand the configuration with minimal application downtime. Unlike the prior single-server configuration, the SAN-Cluster allows moving processing and data serving functions to other nodes while a node is upgraded.

In August 2001, we expanded the TerraServer SAN:

- Added a fourth ESA-10000 storage array (6 TB of "raw" disk, 2.0 TB visible to the O/S)
- Expanded the two existing SAN fabrics to include a third Brocade Silkworm 2800
- Added a separate Brocade Silkworm 2800 to support a third SAN fabric for the tape library
- Added a third HBA controller to each database server
- Added the ADIC Scalar 1000 tape library to the third SAN fabric
- Added the Veritas NetBackup DataCenter utility to the system software configuration

Figure 4 shows these new components (green shading) added to the existing configuration (grey shading).

Figure 4: Cluster Expansion

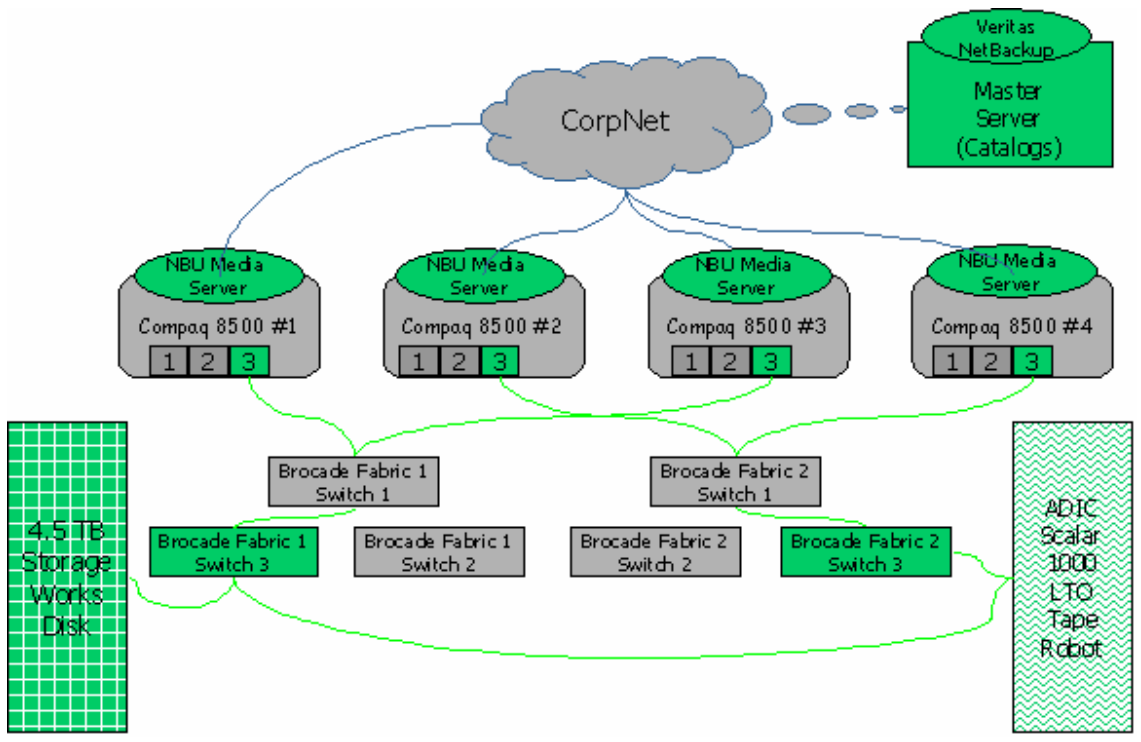

The challenge was to develop an upgrade plan that minimized the database and application downtime while incrementally upgrading the cluster. We were particularly concerned about: (1) adding storage online, and (2) recovery from an operations mistake during the upgrade.

Conventional wisdom was to shutdown the cluster, add the storage, and then re-boot one node at a time. The process that Windows operating system goes through to discover disks is complex. In a cluster environment it is important to orchestrate when each node discovers the new storage volumes, otherwise data corruption is possible. We wanted to perform the volume discovery process without shutting down the cluster.

Given the risk in possible corruption and data-loss, the second major concern was being able to restore lost data from off-line media (tape). As explained earlier, our existing tape backup solution had never worked. For the past year, TerraServer had been operating without any backup solution.

We worked with Compaq and ADIC technicians to develop a plan (see Table 2) that would minimize downtime to end users. Through a series of meetings, we walked through a plan to minimize any mistakes or errors that could cause catastrophic failure or unexpected downtime.

Table 2: Cluster Upgrade Plan and Schedule

| | |
|---|---|
| Day 1<br><br>0 downtime | Install Veritas NetBackup Master Database and Administration component on a remote computer in Redmond. (This tool schedules and maintains a list of the backup operations performed and to be performed.) |
| Day 2<br><br>129 secs downtime | 1) Install new Brocade Silkworm switches (2 on existing fabric, and 1 new fabric);<br>2) Install new HBA (fabric 3) on spare processor not handling active database (SQL4).<br>3) Install NetBackup Media Manager component and driver<br>4) Move active database DRG from SQL1 to spare SQL4 (39 seconds down)<br>5) Install new HBA on SQL1.<br>6) Install NetBackup Media Manager component and driver.<br>7) Move active database DOQ0 from SQL2 to SQL1 (43 seconds down)<br>8) Install new HBA on SQL2<br>9) Install NetBackup Media Manager component and driver<br>10) Move active database DOQ1 from SQL3 to SQL2 (47 seconds down)<br>11) Install new HBA on SQL3<br>12) Install NetBackup Media Manager component and driver |
| Day 2 | Backup DRG database to Scalar 1000 tape library |
| Day 3 | Move tapes from tape library to test lab, restore the database. |
| Day 4 | Run SQL queries to verify the contents of the database. |
| Day 5 | Backup DOQ0 database to Scalar 1000 tape library |
| Day 5 | Move tapes from tape library to test lab, restore the database. |
| Day 6 | Run SQL queries to verify the contents of the database. |
| Day 7 | Backup DOQ1 database to Scalar 1000 library |
| Day 8 | Move tapes from tape library to test lab, restore the database. |
| Day 9 | Run SQL queries to verify the contents of the database. |
| Day 10 | Install 4th ESA-10000 rack, power up, and connect to SAN. Use SAN management software to create three 678GB RAID-10 volumes (triple mirrored, thirty72.8 GB drives each) |
| Day 11<br><br><br>127 secs downtime<br><br><br><br><br>41 secs downtime<br>47 secs downtime<br>44 secs downtime | Upgrade the SAN firmware and SAN S/W on the controllers<br>Add SAN switch to fabric #1 and #2<br>Fail all database resources to the spare node (SQL4)<br>Allow node 1 (SQL1) to discover the three new volumes.<br>Power down SQL1.<br>Allow node 2 (SQL2) to discover the three new volumes.<br>Power down SQL2.<br>Allow node 3 (SQL3) to discover the three new volumes.<br>Power down SQL3.<br>Allow node 4 (SQL4) to discover the three new volumes.<br>Power up SQL1. Failover DRG database from SQL4 to SQL1<br>Power up SQL2. Failover DOQ0 database from SQL4 to SQL2<br>Power up SQL3. Failover DOQ1 database from SQL4 to SQL3.<br>Verify cluster health in Cluster Admin tool |
| Day 12 | Add three drives to cluster resource group.<br>Add drive T to DRG database resources.<br>Add drive U to DOQ0 database resources.<br>Add drive V to DOQ1 database resources. |

We agreed to do the upgrade in three phases:

1. Add the 3rd SAN fabric, Scalar 1000 tape library, and install Veritas NetBackup utility.
2. Backup, restore, and verify each database.
3. Expand the storage by 6 TB adding three 678 GB volumes – one for each active database server.

Table 2 shows the plan and schedule. We were not able to execute the plan in twelve consecutive days. The TerraServer project was a "part-time" job for all members of the team in Microsoft, Compaq, ADIC, and Veritas. We performed the above steps over a period of two months beginning in early August 2001 and worked around each other's work and vacation schedules. Our plan was also interrupted by the tragedy of 9/11.

We planned for the total downtime of 450 seconds during the upgrade. We measure downtime as the amount of time any of the three TerraServer database partitions is unavailable. In reality, only one of the three database partitions is unavailable at any point in time since we failover a database resource to an alternate node. However, we count "partial unavailability" as downtime and do not attempt to amortize it. We estimate that failing a database partition from one node to another to take 50 seconds. Thus, 450 seconds assumes that we needed to failover the database partitions 9 times.

In reality, the database servers can failover from one node to another in less than 50 seconds. The failover time varies with the size of the database, load on the system, and amount of recovery processing required by SQL Server. We have witnessed failovers taking as little as 29 seconds and long as 1.5 minutes. During the cluster expansion process, all failovers ranged between 39 and 47 seconds.

The plan was reviewed by experts within Microsoft, Compaq, ADIC, and Veritas familiar with the intricacies of upgrading clusters and hardware to SANs. Even with a well reviewed plan, we could not and did not anticipate everything that could go wrong. During the Day 1, the installation of the Veritas NetBackup utility required a reboot of the system. We had hoped to accomplish install both the new software and upgrade the HBA firmware at the same time. But the installation process did not allow that. An operational mistake forced a second un-planned re-boot of each system again during Day 1.

The total down-time for the cluster upgrade was 636 seconds (10 minutes, 36 seconds). The maximum downtime allowed in a year to achieve 99.99% availability is 3154 seconds. While the cluster expansion exceeded its 450 second downtime budget by 50%, we were still well within our goal of operating the entire system at 99.99%.

## 3.3 Hardware Failures

The amount of unplanned downtime due to hardware, system software, or operations was minimal. The SAN hardware – disks, controllers, and SAN switches – performed flawlessly for the entire three years. Approximately 24 disk drives failed and were replaced. Two HSG80 controllers failed and were replaced. One SAN switch completely failed and was replaced. Because of the redundancy built into the SAN configuration, there was no measured down-time due to SAN equipment failures. In terms of reliability and performance we highly recommend the Compaq StorageWorks storage solution.

The four Compaq ProLiant 8500, 8-way SMP processors had a handful of failures over the three year period. One processor, SQL1, experienced a number of hardware issues early on which we attributed to burn in. About every 72 to 144 hours the system would have a memory controller error and crash. The processor would automatically recover from the crash and operate normally for another 72 to 144 hours when the cycle would be repeated. We did not notice the failure pattern for a month because the Windows operating system and SQL Server were successfully transferring the database resources to another node in the cluster. The bad news is we had approximately a dozen hardware failures before we noticed the problem. The good news is that this verified and regularly tested the system software's ability to detect and recover from hardware issues.

We had a handful of other component failures such as two HBA controller failures, and one board failure that rendered half of the processors unavailable on one server. SQL1 continued to exhibit occasional hardware failures once the initial memory controller issue was resolved. Our hardware maintenance staff could never diagnose what the random failures were attributed to. They continued to occur sporadically every 60 to 120 days.

## 3.4 Software Failures

The system software (Windows and SQL Server) performed almost flawlessly. No "blue screen" was attributed to software. One database server hung and stopped serving user requests. The operating system and other software on the node had no issue and SQL Server did not report any issue to the system event logs. We luckily noticed the problem after SQL Server was inactive for approximately 50 minutes. Re-booting the server corrected the condition and the hang never occurred again.

We regularly applied operating system and SQL Server patches and software updates. The upgrade process involved moving a database resource from the current node to another node in the cluster, patching the system, and repeating the process on the other three nodes. This procedure worked flawlessly and we experienced 90 seconds to two minutes of downtime. One SQL upgrade experienced corrupted system file. The SQL system failed to restart after the patch was applied. It took a few hours to discover that our upgrade file was corrupt. Replacing the upgrade file and repeating the upgrade process corrected the problem. Fortunately other nodes in the cluster were supporting the database resources, thus we did not experience any additional downtime. This experience verified the value of having an "N+1" cluster where we could tolerate having a processor off-line for maintenance.

## 3.5 Operations Mistakes

We had two major hardware and software maintenance events. The first occurred when we expanded the SAN from 3 racks to 4 racks of equipment, adding seven terabytes of disk and adding a ADIC Scalar 1000 tape robot. This event was discussed in the Cluster Expansion section. The second event occurred when we upgraded the SAN firmware running in the disk controllers and the HBA controllers on the cluster processors.

The SAN-firmware upgrade event caused the longest service outage. We originally planned to take the SAN off-line for 90 minutes to perform the firmware upgrade on the HSG80 controllers, the ADIC Scalar 1000 controller, and the processor HBA controllers. The firmware was upgraded on devices on schedule. When the configuration was re-started, it appeared that all the data on all the disk drives was lost.

The operations staff executed the normal steps to correct the problem, but they were not able to diagnose the trouble. The issue was escalated within Microsoft engineering who discovered that the disk signatures required by the SAN hardware and operating system had been corrupted during the upgrade process. Restoring the correct disk signatures to each volume corrected the problem. Unfortunately, resolving this well-known, but obscure bug in the SAN upgrade process took over 12 hours to discover and the whole process caused a 17.5 hours outage and considerable emotional stress.

Other operations tasks caused downtime addition to hardware and system software maintenance. Generally these were brief planned outages such as moving database resources to particular processors, demonstrating failover events to customers, etc. The most frequent activity was changing the SQL Server software's user account password. The Data Center policy was to change the password every 90 days. Upgrading the SQL Server account password initially required that the SQL Server system be re-started in order for the password change to take effect. This would require all three active databases to be unavailable for approximately 60 seconds while the system was restarted. Later, a program was developed that could update the SQL Server account password without restarting the system. This eliminated any downtime required to reset the account password. However, the program that reset the password had its flaws. If the new password was entered incorrectly, then the SQL Server system would stop serving requests without failing over to another system. The incorrect password update event occurred twice. The first time the problem occurred, we immediately noticed that one server stopped serving requests. However, it took approximately 30 minutes to determine that it was the account password reset that caused the problem. The second time the SQL Server stopped handling requests in the middle of the night. We did not detect and correct the problem until several hours later.

Table 3 summarizes the TerraServer SAN and Cluster downtime. The SAN and Cluster was in service for over 3 years (28,647 hours.)

**Table 3: Summary of outages by category.**

| Downtime Reason | Planned | Unplanned |
|---|---|---|
| **Hardware Failures** | | **0.09 hrs** |
| **Hardware/Firmware Upgrades** | 1.75 hrs | **15.80 hrs** |
| **System Software** | 0.35 hrs | **0.83 hrs** |
| **Operations** | 0.16 hrs | **8.39 hrs** |
| **Downtime** | 2.26 hrs | **25.11 hrs** |
| **Per Cent Available** | 99.99% | **99.91%** |
| **Total Downtime** | **27.37 hrs** | |
| **Per-Cent Available** | **99.90%** | |

In summary, the majority of the downtime was due to maintenance and operational mistakes – a lesson learned many years ago [Gray]. If we ignore downtime due to "operations", the system the goal of four-9's but in reality the system delivered three 9's of availability.

We ran into trouble in two areas: (1) performing complicated tasks for the first time and (2) performing tasks that could not be verified (the account password reset). Our configuration was one of the largest Windows based Compaq StorageWorks SAN and four-node cluster constructed. It was the first Windows and SQL Server cluster our staff and many of the Compaq staff had been exposed to.

In retrospect, we could have eliminated much of the unplanned downtime if we had built an additional small test cluster to train on. This would have eliminated the 16+ hours of downtime we experienced upgrading the SAN firmware, and would have reduced the amount of downtime upgrading the cluster configuration.

# 4 Summary and Conclusion

We operated the TerraServer Cluster-SAN for three years. In that time it delivered several terabytes of images and several billion page-views to end users. The database continually grew with online updates. The hardware and software were quite reliable, and the system was easy to operate when no changes were needed. It demonstrates both a high-traffic web server and a high-traffic web service.

The Cluster-SAN design masked most hardware failures and allowed us to do online hardware and software maintenance. Together the Compaq StorageWorks SAN, Microsoft Cluster Services, and SQL Server 2000 provide a highly reliable and high-performance data processing environment. The environment was easy to operate and the availability features such as transparent resource failover performed flawlessly. We highly recommend our configuration to anyone deploying large SQL Server applications in a high transaction volume environment.

As Table 3 showed, the TerraServer failed to achieve four nines of availability – 99.99%. The vast majority of the downtime was due to operations mistakes that could have been avoided with improved procedures.

Our configuration was large and complex. The operational mistakes occurred when we were doing complex tasks for the very first time, or tasks that could not be verified on our tests systems. Having a geo-plex or a test environment where we could learn how to diagnose the problems we encountered without accruing downtime in the production environment would have been a good investment.